\documentclass[journal]{IEEEtran}

\usepackage{amsmath}
\usepackage{array}
\usepackage{algorithm}
\usepackage{multirow}
\usepackage{graphicx}
\usepackage{epstopdf}
\usepackage{epsfig}
\usepackage{subfigure}
\usepackage{threeparttable}
\usepackage{tabularx}
\usepackage{relsize}
\usepackage{cite}
\usepackage{tikz}
\usepackage{enumerate}

\usepackage[pdftex]{pict2e}
\usepackage{stfloats}
\usepackage{algpseudocode}
\usepackage{csquotes}

\usepackage{amssymb}

\usepackage{cuted}
\usepackage{lipsum}
\DeclareGraphicsExtensions{.pdf, .png, .jpg, .EPS, .tif}

\DeclareMathOperator*{\argmax}{arg\,max}
\newcommand*{\argmaxl}{\argmax\limits}

\begin{document}

\title{Optimal Power Allocation in Uplink NOMA with Simultaneous Cache-Enabled D2D Communications }
    \author{Aditya Powari and Daniel K. C. So, \IEEEmembership{Senior Member, IEEE} 
\vspace{-20pt}

\thanks{
The authors are with the Department of Electrical and Electronic Engineering, University of Manchester, Manchester M13 9PL, United Kingdom (email: adityapowari@gmail.com; d.so@manchester.ac.uk).}

}

\maketitle

\begin{abstract}
Non-orthogonal multiple access (NOMA) is widely viewed as a potential candidate for providing enhanced multiple access in future mobile networks by eliminating the orthogonal distribution of radio resources amongst the users. Nevertheless, the performance of NOMA can be significantly improved by combining it with other sophisticated technologies such as wireless data caching and device-to-device (D2D) communications. In this letter, we propose a novel cellular system model which integrates uplink NOMA with cache based device-to-device (D2D) communications. The proposed system would enable a cellular user to upload data file to base station while simultaneously exchanging useful cache content with another nearby user. We maximize the system sum rate by deriving closed form solutions for optimal power allocation. Simulation results demonstrate the superior performance of our proposed model over other potential combinations of uplink NOMA and D2D communications.
\end{abstract}
\vspace{-0.5cm}
\section{Introduction}

\IEEEPARstart{N}{on-orthogonal} multiple access (NOMA) eliminates the orthogonal sharing of available radio resources amongst the users \cite{Yang_2019}, thereby allowing the users to utilize the available radio resources concurrently. In uplink NOMA, different user equipments (UEs) transmit their messages to the base station (BS) at different transmission powers depending on the UE-BS channel conditions \cite{Cai_2018}. At the BS (receiver), successive interference cancellation (SIC) \cite{Ding_2020} is implemented to decode the messages from these different users. Depending upon the UE-BS channel gains, SIC decodes the user messages from strongest to weakest channel gain. Research in uplink NOMA is predominantly concerned with power control for multi-user messages that would ensure SIC and decoding order preservation at the BS \cite{Imari_2014}. In \cite{Yang_2016}, a dynamic power allocation scheme has been proposed for guaranteeing Quality-of-Service (QoS) to the users in uplink NOMA. Outage and sum rate analysis for a novel uplink power control scheme is shown in \cite{Zhang_2016}. With optimal power allocation, NOMA systems are shown to achieve superior sum rates when compared to OMA techniques \cite{Abbasi_2017}. However, there exists a significant scope for further enhancing its performance by integrating it with sophisticated technologies such as data caching and D2D communications.\\
\indent Wireless data caching is known to improve the efficiency of cellular networks as it enables the storage of popular content on the UEs \cite{Bastug_2014}. This helps in reducing the BS load and system latency \cite{Wang_2017}. In \cite{Wu_2019}, it is shown that when UEs in proximity request popular contents, it is highly probable that the requested contents are located in caches of close-by UEs. The likelihood of this happening is further increased when multiple nearby UEs are each caching different contents  \cite{Ma_2020}. Moreover, due to limited cache memory at the UEs, accurate identification of cache content is necessary \cite{Zhang_2018}. 

\indent In addition, D2D communications can provide an avenue for exchanging useful cache contents between the UEs that are in proximity. When operated in underlay mode, D2D communications ensures higher spectral efficiency in the system at the expense of increased interference \cite{Liu_2012}. This combination of wireless data caching and D2D communications is referred to as cache-enabled D2D communications. Furthermore, unlike D2D communications, the performance of NOMA degrades when the channel gains of multiple users are similar \cite{Kevin_2024}. Thus, integrating uplink NOMA with cache-enabled D2D communications would ensure a reduction in latency while achieving higher spectral efficiency. 

\indent In this letter, we propose a novel uplink NOMA system with simultaneous cache-enabled D2D communications. 
Simultaneous exchange of cache is achieved by operating the UEs on full duplex (FD) transmission mode. Although FD would lead to self-interference (SI) at the UEs, its effect would be mitigated via optimization. We formulate a sum rate maximization problem with QoS constraints and then derive optimal power allocation values in analytical form, which can be implemented efficiently. 
\vspace{-0.3cm}
\section{System Model}
We consider a two-user scenario where each UE is required to upload different data files to the BS via uplink NOMA, while simultaneously exchanging cache content with each other in underlay mode via FD D2D communication. It is imperative that the user pairing technique \cite{Ding_2016} should pair up two UEs that are in proximity and possess useful cache files for one another. In our model, each UE transmits its uplink file and its cache file together as a D2D-NOMA signal to the BS as well as to the paired UE. Depending upon the geographical locations of the UEs within a cell, one UE would have a stronger UE-BS channel gain than the other UE. For simplicity, we denote the UE having stronger UE-BS channel gain as UE\textsubscript{1} while the weaker one is denoted as UE\textsubscript{2}.\\
\begin{figure}[t!]
     \centering
     \includegraphics[width=3.5in,height=2.0in]{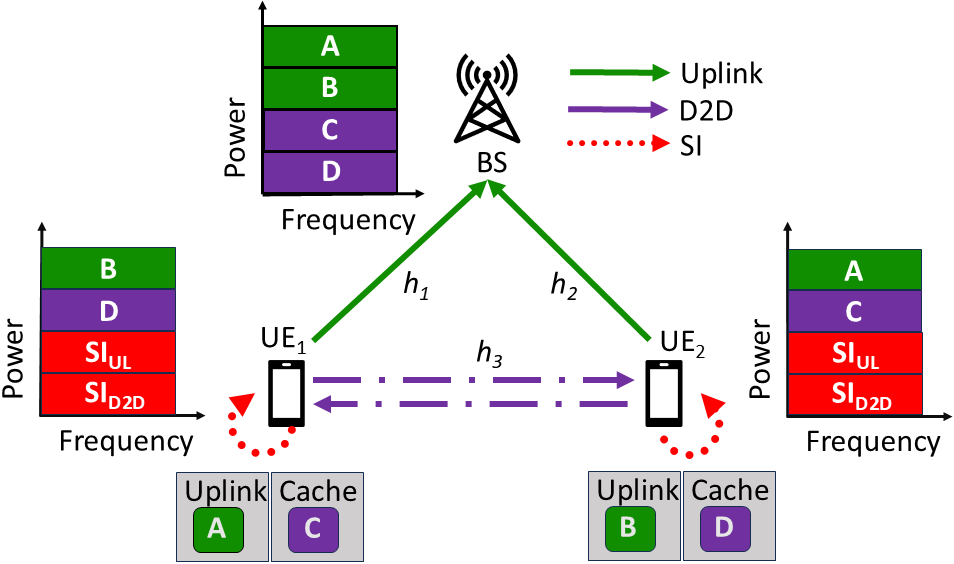}
     \caption{The proposed system model with the received power spectra for each node in descending order}
     \label{system_model}
     \vspace{-0.4cm}
 \end{figure}
\indent The proposed system model is shown in Fig. \ref{system_model} where UE\textsubscript{1} and UE\textsubscript{2} are required to upload data files A and B to the BS respectively. Files C and D are the cached files which are stored in the cache memories of UE\textsubscript{1} and UE\textsubscript{2} respectively. Thus, UE\textsubscript{1} and UE\textsubscript{2} transmit files (A, C) and (B, D) as a D2D-NOMA signal respectively. Since cache files are the popularly requested contents, we assume that they are also available in the BS cache. As a result, cache-enabled interference cancellation (CIC) \cite{Xiang_2019} is implemented at the BS to mitigate interference from files C and D. For brevity, the overall channel-to-noise ratio (CNR) is denoted by $|h_i|^2=\frac{\delta_i|H_i|^2}{L_p^i BN_0}$, with $i\in\{1,2,3\}$ referring to the UE\textsubscript{1}-BS, UE\textsubscript{2}-BS and the UE\textsubscript{1}-UE\textsubscript{2} links respectively. Here, $\delta_i$ is the lognormal shadowing, $|H_i|^2$ is the fading channel gain, $L_p^i$ is the exponential path loss, $B$ is the total bandwidth and $N_0$ is the noise power spectral density. The channel gains $|H_1|^2$, $|H_2|^2$ are modelled as Rayleigh fading while $|H_3|^2$ as Rician fading. Furthermore, SI is observed at both the UEs due to their FD mode. The CNR of the SI channel is denoted by $|h_{SI}|^2=\frac{|H_{SI}|^2}{\zeta L_{p}^{SI} BN_0}$, where $|H_{SI}|^2$ is the Rician fading gain, $L_{p}^{SI}$ is the free space path loss and $\zeta$ is the SI cancellation factor. It must be noted that there are two SI files at each UE, one due to uplink file (SI\textsubscript{UL}) and the other due to cache file (SI\textsubscript{D2D}).\\
\indent Let $P_{UE}$ be the available transmission power at the UEs, $\alpha_1$ and $\alpha_2$ be the power allocation factors for UE\textsubscript{1} and UE\textsubscript{2} respectively. In conventional uplink NOMA the entire $P_{UE}$ is used by each UE for optimal performance. 
We now preserve the decoding order of the received files shown in the power spectrum of BS and both the UEs as seen in Fig. \ref{system_model}. The domain of both UE power allocation factors is $\alpha_n\in[0,0.5]\; \forall \; n\in\{1,2\}$. The major share of the available UE transmission power, i.e., $(1-\alpha_n)P_{UE}$ is allocated for uplink NOMA message while the remaining power, i.e.,  $\alpha_nP_{UE}$ is allocated for D2D communications. To preserve the decoding order at BS we need to ensure that file A (from the strong user) is received at a higher power than file B (from the weak user). This requires,
\begin{equation}
    \alpha_1 < \frac{(|h_1|^2-|h_2|^2) + \alpha_2 |h_2|^2}{|h_1|^2},
    \label{alpha_1_upper_1}
\end{equation}
to hold true, where $|h_1|^2>|h_2|^2$. Considering (\ref{alpha_1_upper_1}) along with the domain of $\alpha_1$, an upperbound on $\alpha_1$ can be formulated as
\begin{equation}
    \overline{\alpha_1} = \min \bigg\{\frac{(|h_1|^2-|h_2|^2) + \alpha_2 |h_2|^2}{|h_1|^2},0.5 \bigg\}.
    \label{alpha_1_upperbound}
\end{equation}
On the other hand, the decoding order at both the UEs has already been ensured because more power is allocated to uplink transmissions at both UEs, and hence SI\textsubscript{UL} is always larger than SI \textsubscript{D2D}. Therefore, by ensuring that the cache file is received at a higher power than SI\textsubscript{UL} file at both UEs would guarantee order preservation. Mathematically, this would be possible if
\vspace{-0.2cm}
\begin{subequations}
\begin{equation}
        \alpha_1 > 1 - \frac{\alpha_2 |h_3|^2}{|h_{SI}|^2}
        \label{alpha_1_lower_1}
    \end{equation}
\text{and}
\begin{equation}
        \alpha_1 > \frac{(1-\alpha_2) |h_{SI}|^2}{|h_3|^2}
        \label{alpha_1_lower_2}
    \end{equation}
    \label{alpha_1_lowerbound}
 \end{subequations}\\
hold true simultaneously. Thus, by considering (\ref{alpha_1_lowerbound}) and the domain of $\alpha_1$, we can define a lowerbound on $\alpha_1$ as
 \begin{equation}
     \underline{\alpha_1} = \Bigg[ \max\bigg\{ 1 - \frac{\alpha_2 |h_3|^2}{|h_{SI}|^2}, \frac{(1-\alpha_2) |h_{SI}|^2}{|h_3|^2} \bigg\} \Bigg]_{0}^{0.5}.
 \end{equation}
 Moreover, the received Signal-to-Interference-plus-Noise ratios (SINRs) after SIC and CIC for receiver $i$, receiving file $j$, denoted as $\gamma_{i\rightarrow j}$, are:
    \vspace{-0.3cm}
    \begin{equation}
        \gamma_{BS\rightarrow A} = \frac{|h_1|^2 (1-\alpha_1)P_{UE}}{|h_2|^2 (1-\alpha_2) P_{UE} + 1}
        \label{gamma_BS_A}
    \end{equation}
    \begin{equation}
        \gamma_{BS\rightarrow B} = |h_2|^2 (1-\alpha_2) P_{UE} 
        \label{gamma_BS_B}
    \end{equation}

    \begin{equation}
        \gamma_{UE_{1}\rightarrow B} = \frac{|h_3|^2 (1-\alpha_2) P_{UE}}{|h_3|^2 \alpha_2 P_{UE}+|h_{SI}|^2 P_{UE} + 1}
         \label{gamma_UE1_B}
    \end{equation}
    \begin{equation}
        \gamma_{UE_{1}\rightarrow D} = \frac{|h_3|^2 \alpha_2 P_{UE}}{|h_{SI}|^2 P_{UE} + 1}
         \label{gamma_UE1_C}
    \end{equation}
    
    \begin{equation}
        \gamma_{UE_{2}\rightarrow A} = \frac{|h_3|^2 (1-\alpha_1) P_{UE}}{|h_3|^2 \alpha_1 P_{UE}+|h_{SI}|^2 P_{UE} + 1} 
         \label{gamma_UE2_A}
    \end{equation}
    \begin{equation}
        \gamma_{UE_{2}\rightarrow C} = \frac{|h_3|^2 \alpha_1 P_{UE}}{|h_{SI}|^2 P_{UE} + 1} 
         \label{gamma_UE2_D}
    \end{equation}

The resulting maximum achievable data rates for each received file can be computed using $R_{i\rightarrow j} = B \log_2 (1+ \gamma_{i\rightarrow j})$. The system sum rate ($R_{sum}$) is the sum of uplink rates and the D2D communication rates, which is defined as
\vspace{-0.1cm}
\begin{equation}
    R_{sum} = R_{UL} + R_{D2D}, 
    \label{sum_rate}
\end{equation}
where $R_{UL}= R_{BS\rightarrow A} + R_{BS\rightarrow B}$, represents the total uplink NOMA rate and $R_{D2D}=R_{UE_{1}\rightarrow D} + R_{UE_{2}\rightarrow C}$, represents the total D2D rate.

\section{Optimization Problem and Solution}
The objective is to maximize $R_{sum}$ by determining the optimal values of UE power allocation factors, i.e., $\alpha_1$ and $\alpha_2$, subject to several constraints as follows:
\begin{subequations}
\begin{alignat}{2} 
\!\max_{\alpha_1,\alpha_2} \quad & R_{sum} \label{eq:a}\\
\textrm{s.t.} \quad & R_{BS\rightarrow A} \geq R_{min-A} \label{eq:b}\\
    \quad & R_{BS\rightarrow B} \geq R_{min-B} \label{eq:c} \\
    \quad & R_{UE_1\rightarrow D} \geq R_{min-D} \label{eq:d} \\
    \quad & R_{UE_2\rightarrow C} \geq R_{min-C} \label{eq:e} \\
    \quad & R_{UE_2\rightarrow A} \geq R_{min-A} \label{eq:f}\\
    \quad & R_{UE_1\rightarrow B} \geq R_{min-B} \label{eq:g}\\
    \quad & \underline{\alpha_1} \leq \alpha_1 \leq \overline{\alpha_1} \label{eq:h}\\
    \quad & 0 \leq \alpha_2 \leq 0.5 \label{eq:i} 
\vspace{-0.3cm} 
\end{alignat}

\end{subequations}

{\parindent0pt
    where $R_{min-j}$ is the minimum data rate required by the file $j$. The constraints (\ref{eq:b}) to (\ref{eq:e}) are the minimum file rates required to maintain QoS levels, (\ref{eq:f}) and (\ref{eq:g}) ensures successful SIC at the UEs, (\ref{eq:h}) is the modified domain of $\alpha_1$ necessary to ensure decoding order preservation and (\ref{eq:i}) is the domain of $\alpha_2$. Furthermore, $\gamma_j$ represents the minimum required SINR to ensure QoS for each file $j$ and is calculated as $\gamma_j=2^\frac{R_{min-j}}{B} -1$. \\
}
\indent The double derivative of $R_{sum}$ w.r.t $\alpha_1$ is
\begin{equation}
\frac{d^2 R_{sum}}{d \alpha_1 ^2} = -\frac{B {P_{UE}}^2}{ln(2)}\bigg[\frac{|h_3|^2}{\chi_1^2} + \frac{|h_1|^2}{\chi_2^2} \bigg] ,
    \label{alpha_1_d2}
\end{equation}
where $\chi_1= 1 + |h_{SI}|^2 P_{UE} + |h_3|^2 \alpha_1 P_{UE} $,\\
$\chi_2= 1 + |h_1|^2 P_{UE} ( 1 - \alpha_1) + |h_2|^2 P_{UE} (1-\alpha_2) $.\\
\indent From (\ref{alpha_1_d2}), it is clear that the double derivative is unconditionally negative. Thus, $R_{sum}$ is concave w.r.t $\alpha_1$. Moreover, it is also evident that (\ref{alpha_1_d2}) contains no points of discontinuities. Therefore, the optimal $\alpha_1$ value can be obtained by examining the behaviour of its first derivative w.r.t the given constraints. The first derivative of $R_{sum}$ w.r.t $\alpha_1$ is
\begin{equation}
\frac{d R_{sum}}{d \alpha_1} = \frac{B P_{UE} \phi_1}{ln(2)  [1+|h_{SI}|^2 P_{UE}+|h_3|^2 \alpha_1 P_{UE}] \phi_2} ,
    \label{alpha_1_d1}
\end{equation}
where $\phi_1=|h_1|^2|h_3|^2P_{UE}(1-2\alpha_1) + |h_2|^2|h_3|^2P_{UE}(1-\alpha_2) + |h_3|^2 - |h_1|^2 - |h_1|^2|h_{SI}|^2P_{UE} $ and \\ 
$\phi_2 = 1 + |h_1|^2P_{UE}(1-\alpha_1) + |h_2|^2 P_{UE}(1-\alpha_2) $.\\ 
Thus, $R_{sum}$ is a monotonically increasing function w.r.t $\alpha_1$ as (\ref{alpha_1_d1}) is always positive. As a result, a higher $\alpha_1$ value would yield a higher $R_{sum}$. However, increasing $\alpha_1$ would reduce (\ref{gamma_BS_A}) and (\ref{gamma_UE2_A}), which would in turn reduces their data rates and violate constraints (\ref{eq:b}) and (\ref{eq:f}). Furthermore, it can be comprehended that $ \gamma_{UE_{2}\rightarrow A}>  \gamma_{BS\rightarrow A}$ because $\gamma_{UE_{2}\rightarrow A}$ utilizes the strongest channel (the D2D one) and experiences comparatively less interference from the low powered cache file and suppressed SI, in contrast to $\gamma_{BS\rightarrow A}$. Hence, satisfying constraint (\ref{eq:b}) would ensure constraint (\ref{eq:f}) holds true. Due to the monotonically increasing nature of $\alpha_1$, we can obtain its value by solving $\gamma_{BS\rightarrow A} = \gamma_A$. Moreover, constraint (\ref{eq:h}) imposes a domain constraint on $\alpha_1$ as well. Thus, by incorporating (\ref{eq:h}), the optimal $\alpha_1$ value is  
\begin{equation}
    \alpha^*_1 = \Bigg[ \frac{|h_1|^2 P_{UE} - \gamma_A \big(|h_2|^2 P_{UE}(1-\alpha_2)+1 \big)}{|h_1|^2 P_{UE}}  \Bigg]_{\underline{\alpha_1}}^{\overline{\alpha_1}} . 
    \label{alpha_1_optimal}
\end{equation}
\indent By substituting (\ref{alpha_1_optimal}) into (\ref{sum_rate}), we establish $R_{sum}$ as a pure function of $\alpha_2$, which can now be optimized. Similar to the above, the double derivative is
\begin{equation}
\frac{d^2 R_{sum}}{d \alpha_2 ^2} = -\frac{B {P_{UE}}^2}{ln(2)}\Bigg[ \frac{{\gamma_A}^2 |h_2|^2 |h_3|^2}{\psi_1^2} + \frac{|h_2|^2}{\psi_2^2} + \frac{|h_3|^2}{\psi_3^2}\Bigg],
\label{alpha_2_d2}    
\end{equation}
where $\psi_1=|h_1|^2(1+|h_3|^2P_{UE}+|h_{SI}|^2P_{UE}) - \gamma_A |h_3|^2 (1+|h_2|^2 P_{UE} (1-\alpha_2)) $, $\psi_2 = 1+|h_2|^2P_{UE}(1-\alpha_2)$ and $\psi_3 = 1+|h_{SI}|^2 P_{UE} + |h_3|^2 \alpha_2 P_{UE}$. Thus, $R_{sum}$ is always concave w.r.t $\alpha_2$ as (\ref{alpha_2_d2}) is unconditionally negative. However, (\ref{alpha_2_d2}) contains a point of discontinuity which would make convex optimization infeasible. The first term in (\ref{alpha_2_d2}) becomes undefined when 
\begin{equation}
    \alpha_2 = \frac{\gamma_A |h_3|^2 + \Upsilon - |h_1|^2 P_{UE}(|h_{SI}|^2+ |h_3|^2) - |h_1|^2}{\Upsilon} ,
    \label{alpha_2_discontinue}
\end{equation}
{\parindent0pt where $\Upsilon=\gamma_A |h_2|^2 |h_3|^2 P_{UE}$. As this value of $\alpha_2$ is feasible, we cannot implement convex optimization. Therefore, we shorten the search for the feasible optimal value by formulating an upper bound ($\overline{\alpha_2}$) and a lower bound ($\underline{\alpha_2}$) on $\alpha_2$ based on the given constraints. Following this, we derive the stationary point ($\Hat{\alpha_2}$) and then present a case-wise solution for the optimal $\alpha_2$ value. In a nutshell, as $R_{sum}$ is concave w.r.t $\alpha_2$, so if $\Hat{\alpha_2}$ is located in between the bounds, then it is the optimal $\alpha_2$ value. Otherwise, the bound which yields the maximum $R_{sum}$ is the optimal $\alpha_2$ value.}

\indent It can be observed from (\ref{gamma_BS_B}) and (\ref{gamma_UE1_B}) that increasing $\alpha_2$ would decrease $ \gamma_{BS\rightarrow B}$ and $ \gamma_{UE_{1}\rightarrow B}$ respectively, which would violate constraints (\ref{eq:c}) and (\ref{eq:g}). It can be noted that $\gamma_{BS\rightarrow B}>  \gamma_{UE_{1}\rightarrow B}$, as $ \gamma_{BS\rightarrow B}$ is decoded without any interference due to CIC and SIC. Hence, satisfying constraint (\ref{eq:g}) will ensure constraint (\ref{eq:c}) is met.
As $\alpha_2$ increases, constraint (\ref{eq:g}) will limit its maximum value and so would act as an upper bound on $\alpha_2$. This upper bound is obtained by solving $\gamma_{UE_1\rightarrow B}=\gamma_B$ for $\alpha_2$ and by incorporating (\ref{eq:i}) to get
\begin{equation}
    \overline{\alpha_2} = \Bigg[ \frac{|h_3|^2P_{UE} - \gamma_B(|h_{SI}|^2 P_{UE}+1)}{|h_3|^2 P_{UE}(1+\gamma_B)} \Bigg]_{0}^{0.5}.
    \label{alpha_2_upper}
\end{equation}
\indent On the other hand, when we decrease $\alpha_2$, it is observed from (\ref{gamma_BS_A}) and (\ref{gamma_UE1_C}) that both $\gamma_{BS\rightarrow A}$ and $\gamma_{UE_1\rightarrow D}$ would decrease. Notably, the use of $\alpha^*_1$ would ensure QoS for file A. Therefore, we only need to achieve the QoS for file D by satisfying constraint (\ref{eq:d}). Thus, (\ref{eq:d}) would limit the minimum value of $\alpha_2$, in which case the lower bound can be obtained by solving $\gamma_{UE_1\rightarrow D}=\gamma_D$. Moreover, by incorporating the domain constraint (\ref{eq:i}), we can express the lower bound as
\begin{equation}
    \underline{\alpha_2} = \Bigg[\frac{\gamma_D(|h_{SI}|^2 P_{UE}+1)}{|h_3|^2 P_{UE}}\Bigg]_{0}^{0.5} .
    \label{alpha_2_lower}
\end{equation}
\indent By equating the first derivative of $R_{sum}$ w.r.t $\alpha_2$ to zero, we obtain two stationary points
\begin{equation}
     \Hat{\alpha_2} = \frac{1}{\xi_1} \bigg(-\xi_2 \pm \sqrt{{\xi_2}^2 -\xi_3} \bigg) ,
      \label{alpha_2_sp}
\end{equation}
where  $\xi_1 = 6\gamma_A |h_2|^4 |h_3|^4 {P_{UE}}^2$,\\ $\xi_2= 2|h_1|^2|h_2|^2|h_3|^2P_{UE}(1+|h_3|^2P_{UE}+|h_{SI}|^2P_{UE}) + 2\gamma_A |h_2|^4|h_3|^2P_{UE}(1+|h_{SI}|^2P_{UE}) - 4\gamma_A|h_2|^2|h_3|^4P_{UE}(1+|h_2|^2P_{UE})$, and $\xi_3= 12\gamma_A|h_2|^4|h_3|^4 {P_{UE}}^2 \big[ \gamma_A |h_3|^4 + \gamma_A |h_2|^2|h_3|^2(|h_2|^2|h_3|^2{P_{UE}}^2 -2) + |h_1|^2|h_2|^2(1 + 2|h_{SI}|^2P_{UE} -|h_3|^4{P_{UE}}^2 + |h_{SI}|^4{P_{UE}}^2) - |h_1|^2|h_3|^2(1+ |h_3|^2P_{UE}+|h_{SI}|^2P_{UE}) -2\gamma_A|h_2|^2|h_3|^2P_{UE}(|h_2|^2-|h_3|^2+|h_{SI}|^2+|h_2|^2|h_{SI}|^2P_{UE}) \big] $
. Out of these two stationary points, we choose the one that satisfies constraint (\ref{eq:i}) as $\Hat{\alpha_2}$. Thus, the optimal $\alpha_2$ value can be expressed as
\begin{equation}
    \alpha^*_2 = \begin{cases}
        \Hat{\alpha_2} & \text{if }  \Hat{\alpha_2} \in[ \underline{\alpha_2}, \overline{\alpha_2}]\\
        \argmaxl_{\alpha_2\in\{\underline{\alpha_2},\overline{\alpha_2} \}} R_{sum} & \text{otherwise}.
    \end{cases}
    \label{alpha_2_optimal}
\end{equation}


\section{Simulation Results}

As there is no comparable existing work, we compare the performance of the proposed Simultaneous uplink-D2D model with two possible combinations between uplink NOMA and cache-enabled D2D communications, namely the Phased and the Slotted approach. In the Phased approach, the transmissions are divided into two separate phases: the UE uplink transmission phase and the FD D2D cache exchange phase. Due to the two separate phases, the UEs are able to utilize the entire $P_{UE}$ for their transmissions in each phase. On the other hand, the Slotted approach divides the time into two separate slots: one for UE\textsubscript{1} to transmit a D2D-NOMA signal to both the BS and UE\textsubscript{2}, and the other for UE\textsubscript{2} to transmit. Here, power allocation is implemented at both UEs by considering uplink rate constraints for maintaining QoS. In the simulations, the paired UEs are distributed randomly around the BS such that they always remain within a maximum D2D separation with each other. This ensures that the cache-enabled D2D communications remain feasible. The simulation parameters used are stated in Table \ref{simulation_paramters_table}.


 
\begin{table}[h!]
     \centering
     \caption{Simulation Parameters}
     \label{simulation_paramters_table}
     \scalebox{1.1}{
     \begin{tabular}{|c|c|}
        \hline
         \textbf{Parameter} & \textbf{Value} \\
         \hline
         \hline
         \textit{Cell radius} & $250$ \textit{m} \\
         \hline
         \textit{Max D2D separation} & $20$ \textit{m} \\
         \hline
         \textit{Max UE transmission power:} $P_{UE-max}$ & $25$ \textit{dBm} \\
         \hline
        \textit{Total system bandwidth} & $5$ \textit{MHz} \\
         \hline
         \textit{Carrier frequency} & $2$ \textit{GHz} \\
         \hline
         \textit{Shadowing standard deviation:} $\sigma$ & $8$ \textit{dB}\\
         \hline
         \textit{Noise power spectral density:} $N_0$ &  $-174$ \textit{dBm/Hz}\\
         \hline
         \textit{Antenna separation at UE} & $0.3$ \textit{m}\\
         \hline
         \textit{Path loss exponent} & $3$ \\
         \hline
         \textit{SI cancellation factor:} $\zeta$ & $80$ \textit{dB}\\
         \hline
         \textit{$R_{min-j}$ for all data files } & $5$ \textit{Mbps}\\
         \hline
         
     \end{tabular}}    
 \end{table}
 
 \begin{figure}[h!]
     \centering
     \includegraphics[width=3.8in,height=2.0in]{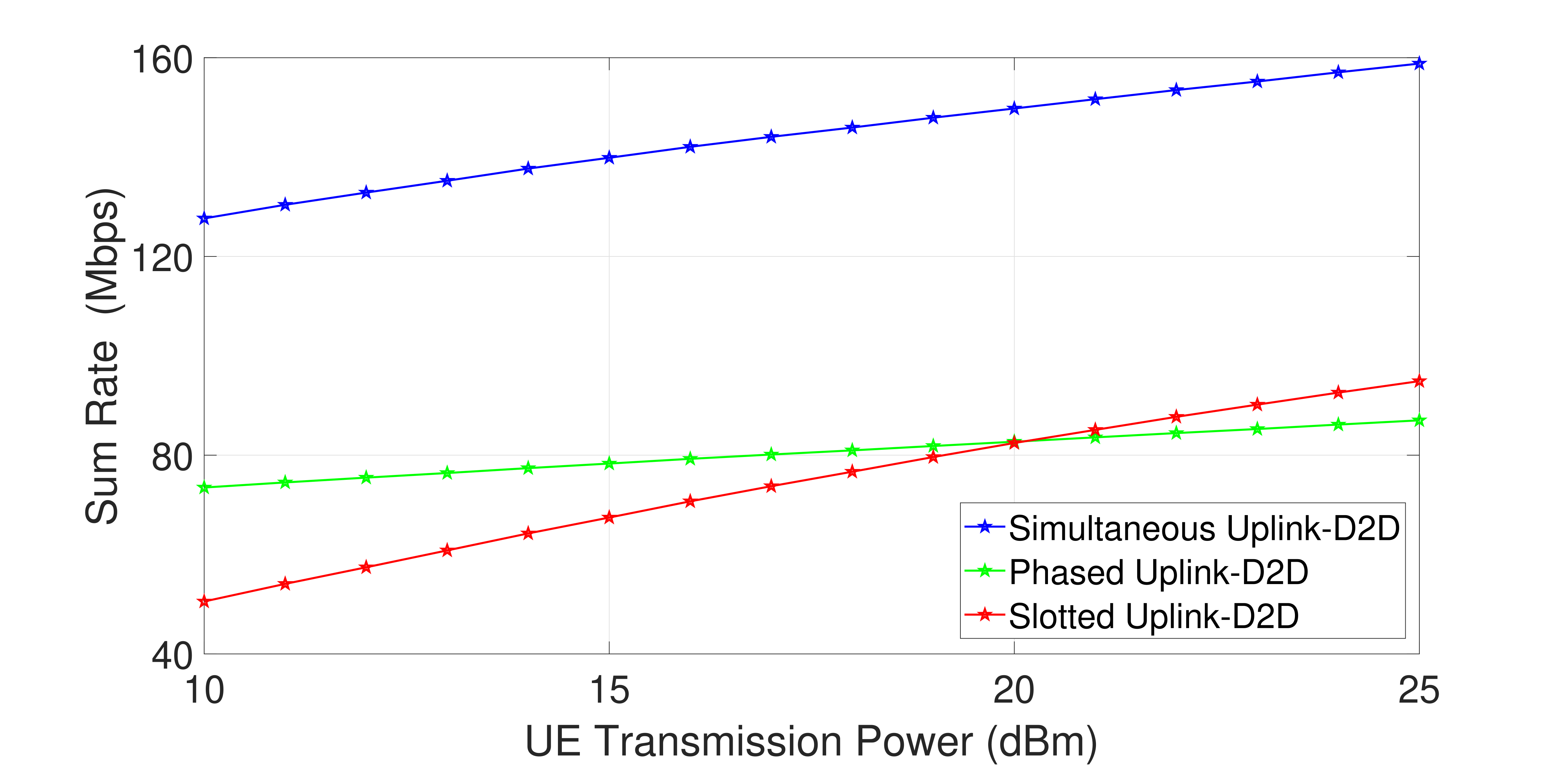}
     \caption{Sum Rate verses UE transmission power}
     \label{sum_rate_figure}
 \end{figure}
 Fig. \ref{sum_rate_figure} shows that as UE transmission power increases, the sum rate increases for all the systems under consideration. This is because of the improvement in the SINRs of the uplink and cache files due to increasing $P_{UE}$. Moreover, it is apparent that the proposed system yields a substantially higher sum rate when compared to the phased and slotted approaches. Hence, coupling simultaneous D2D communications with uplink NOMA is far more beneficial for the system rather than combining them via the phased or slotted approaches.
 
 \begin{figure}[t!]
     \centering
     \includegraphics[width=3.8in,height=2.0in]{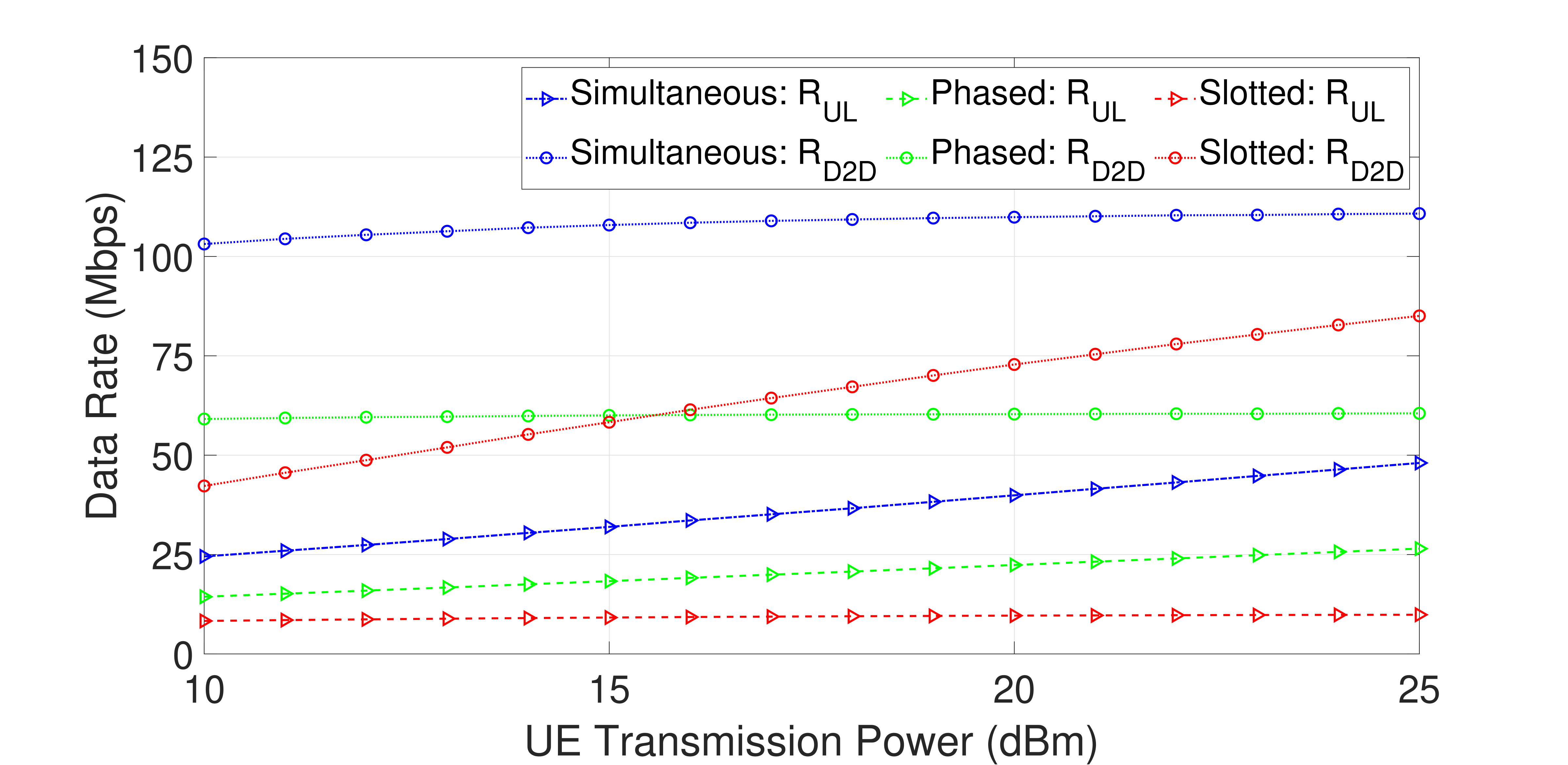}
     \caption{Uplink and D2D rates verses UE transmission power}
     \label{user_rate_figure}
     \vspace{-0.4cm}
 \end{figure}
 Fig. \ref{user_rate_figure} presents the achievable uplink rates ($R_{UL}$) and the achievable D2D rates ($R_{D2D}$) for all systems against UE transmission power. It is evident that the uplink rate achieved from the proposed system supersedes the uplink rates achieved from the phased and slotted approaches. The uplink rate of the phased approach shows a gradual improvement when $P_{UE}$ is increased because it also increases the interference to the strong user signal. On the other hand, power allocation in the slotted approach prioritizes the uplink file QoS after which the remaining power is allocated to the cache file. As a result, the slotted approach yields a constant uplink rate. On the D2D rates, the proposed system yields the highest data rate which highlights the significance of optimal power allocation factors (\ref{alpha_1_optimal}) and (\ref{alpha_2_optimal}). The D2D rate of the phased approach shows a negligible improvement when $P_{UE}$ is increased as it also increases the SI at both the UEs during the D2D phase. Meanwhile, the D2D rate of the slotted approach increases with $P_{UE}$ because the amount of power allocated to the cache file increases with $P_{UE}$. \\
\vspace{-0.6cm}
\begin{figure}[h!]
     \centering
     \includegraphics[width=3.8in,height=2.0in]{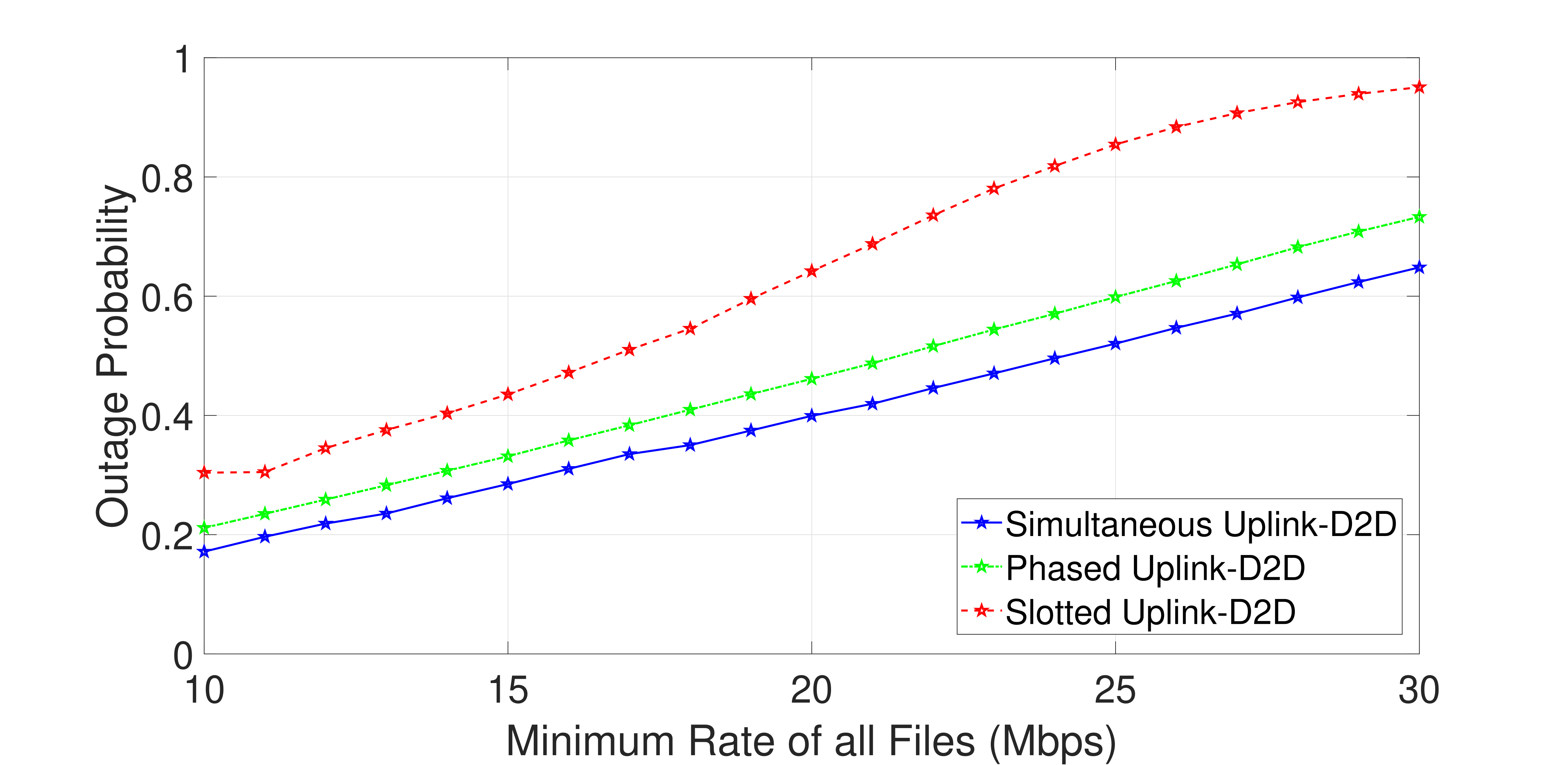}
     \caption{Outage probability against $R_{min-j}$ at $P_{UE-max}$}
     \label{SI_cancellation_figure}
     \vspace{-0.3cm}
 \end{figure}
 \newline
\indent Fig. \ref{SI_cancellation_figure} presents the effect of varying the minimum rates of all the files, i.e., $R_{min-j}\: \forall \: j\in\{A,B,C,D\}$ on the outage probability with the UE transmission power kept fixed at $P_{UE-max}$. It is evident that the proposed system outperforms the phased and the slotted systems over the entire range of $R_{min-j}$. The phased approach performs better than the slotted approach, as it utilizes the entire available $P_{UE}$ to support the high data rate requirements as $R_{min-j}$ increases. On the other hand, the slotted approach shows a steep rise in the outage probability as $R_{min-j}$ increases. The reason behind this behaviour is that power allocated to the uplink files increases with $R_{min-j}$ in order to meet their necessary QoS. As a result, less power is available for the cache files due to which they fail to meet their QoS, leading to outage.\\
\section{Conclusions}
In this letter, we proposed a novel system that integrates uplink NOMA with simultaneous cache-enabled D2D
communications between the UEs. Furthermore, closed-form expressions for the UE power allocation factors were derived to achieve maximum sum rate while adhering to the necessary QoS constraints. Simulation results highlighted the substantial performance gains of the proposed system over the phased and slotted approaches.

\bibliographystyle{IEEEtran}
\bibliography{IEEEabrv,main}

\begin{thebibliography}{10}
\providecommand{\url}[1]{#1}
\csname url@samestyle\endcsname
\providecommand{\newblock}{\relax}
\providecommand{\bibinfo}[2]{#2}
\providecommand{\BIBentrySTDinterwordspacing}{\spaceskip=0pt\relax}
\providecommand{\BIBentryALTinterwordstretchfactor}{4}
\providecommand{\BIBentryALTinterwordspacing}{\spaceskip=\fontdimen2\font plus
\BIBentryALTinterwordstretchfactor\fontdimen3\font minus \fontdimen4\font\relax}
\providecommand{\BIBforeignlanguage}[2]{{%
\expandafter\ifx\csname l@#1\endcsname\relax
\typeout{** WARNING: IEEEtran.bst: No hyphenation pattern has been}%
\typeout{** loaded for the language `#1'. Using the pattern for}%
\typeout{** the default language instead.}%
\else
\language=\csname l@#1\endcsname
\fi
#2}}
\providecommand{\BIBdecl}{\relax}
\BIBdecl

\bibitem{Yang_2019}
K.~Yang, N.~Yang, N.~Ye, M.~Jia, Z.~Gao, and R.~Fan, ``Non-orthogonal multiple access: Achieving sustainable future radio access,'' \emph{IEEE Communications Magazine}, vol.~57, no.~2, pp. 116--121, 2019.

\bibitem{Cai_2018}
Y.~Cai, Z.~Qin, F.~Cui, G.~Y. Li, and J.~A. McCann, ``Modulation and multiple access for 5g networks,'' \emph{IEEE Communications Surveys \& Tutorials}, vol.~20, no.~1, pp. 629--646, 2018.

\bibitem{Ding_2020}
Z.~Ding, R.~Schober, and H.~V. Poor, ``Unveiling the importance of {SIC} in {NOMA} systems—part 1: State of the art and recent findings,'' \emph{IEEE Communications Letters}, vol.~24, no.~11, pp. 2373--2377, 2020.

\bibitem{Imari_2014}
M.~Al-Imari, P.~Xiao, M.~A. Imran, and R.~Tafazolli, ``Uplink non-orthogonal multiple access for {5G} wireless networks,'' in \emph{2014 11th International Symposium on Wireless Communications Systems (ISWCS)}, 2014, pp. 781--785.

\bibitem{Yang_2016}
Z.~Yang, Z.~Ding, P.~Fan, and N.~Al-Dhahir, ``A general power allocation scheme to guarantee quality of service in downlink and uplink noma systems,'' \emph{IEEE Transactions on Wireless Communications}, vol.~15, no.~11, pp. 7244--7257, 2016.

\bibitem{Zhang_2016}
N.~Zhang, J.~Wang, G.~Kang, and Y.~Liu, ``Uplink nonorthogonal multiple access in {5G} systems,'' \emph{IEEE Communications Letters}, vol.~20, no.~3, pp. 458--461, 2016.

\bibitem{Abbasi_2017}
Z.~Q. Al-Abbasi and D.~K.~C. So, ``Resource allocation in non-orthogonal and hybrid multiple access system with proportional rate constraint,'' \emph{IEEE Transactions on Wireless Communications}, vol.~16, no.~10, pp. 6309--6320, 2017.

\bibitem{Bastug_2014}
E.~Bastug, M.~Bennis, and M.~Debbah, ``Living on the edge: The role of proactive caching in {5G} wireless networks,'' \emph{IEEE Communications Magazine}, vol.~52, no.~8, pp. 82--89, 2014.

\bibitem{Wang_2017}
W.~Wang, R.~Lan, J.~Gu, A.~Huang, H.~Shan, and Z.~Zhang, ``Edge caching at base stations with device-to-device offloading,'' \emph{IEEE Access}, vol.~5, pp. 6399--6410, 2017.

\bibitem{Wu_2019}
D.~Wu, Q.~Liu, H.~Wang, Q.~Yang, and R.~Wang, ``Cache less for more: Exploiting cooperative video caching and delivery in {D2D} communications,'' \emph{IEEE Transactions on Multimedia}, vol.~21, no.~7, pp. 1788--1798, 2019.

\bibitem{Ma_2020}
Z.~Ma, N.~Nuermaimaiti, H.~Zhang, H.~Zhou, and A.~Nallanathan, ``Deployment model and performance analysis of clustered {D2D} caching networks under cluster-centric caching strategy,'' \emph{IEEE Transactions on Communications}, vol.~68, no.~8, pp. 4933--4945, 2020.

\bibitem{Zhang_2018}
S.~Zhang, P.~He, K.~Suto, P.~Yang, L.~Zhao, and X.~Shen, ``Cooperative edge caching in user-centric clustered mobile networks,'' \emph{IEEE Transactions on Mobile Computing}, vol.~17, no.~8, pp. 1791--1805, 2018.

\bibitem{Liu_2012}
Z.~Liu, T.~Peng, S.~Xiang, and W.~Wang, ``Mode selection for device-to-device ({D2D}) communication under {LTE}-advanced networks,'' in \emph{2012 IEEE International Conference on Communications (ICC)}, 2012, pp. 5563--5567.

\bibitem{Kevin_2024}
K.~Z. Shen, D.~K.~C. So, J.~Tang, and Z.~Ding, ``Power allocation for {NOMA} with cache-aided {D2D} communication,'' \emph{IEEE Transactions on Wireless Communications}, vol.~23, no.~1, pp. 529--542, 2024.

\bibitem{Ding_2016}
Z.~Ding, P.~Fan, and H.~V. Poor, ``Impact of user pairing on {5G} nonorthogonal multiple-access downlink transmissions,'' \emph{IEEE Transactions on Vehicular Technology}, vol.~65, no.~8, pp. 6010--6023, 2016.

\bibitem{Xiang_2019}
L.~Xiang, D.~W.~K. Ng, X.~Ge, Z.~Ding, V.~W.~S. Wong, and R.~Schober, ``Cache-aided non-orthogonal multiple access: The two-user case,'' \emph{IEEE Journal of Selected Topics in Signal Processing}, vol.~13, no.~3, pp. 436--451, 2019.

\end{thebibliography}
\end{document}